\documentclass[prx,aps,twocolumn,superscriptaddress]{revtex4-2}

\usepackage[pdftex]{graphicx}
\usepackage{amsbsy,amssymb,amsmath,bm,mathtools}
\usepackage{amsfonts}
\usepackage{makecell}
\usepackage{mathrsfs}
\usepackage{amssymb}
\usepackage{euscript}
\usepackage{xspace}
\usepackage{bm}
\usepackage{color}
\usepackage{url}
\usepackage[section]{placeins}
\usepackage[colorlinks=true,linkcolor=blue,citecolor=blue]{hyperref}

\begin{document}


\title{Point-Group Equivariant Neural Networks for Local Dynamics in Lattice Systems}

\author{Yunhao Fan}
\affiliation{Department of Physics, University of Virginia, Charlottesville, VA 22904, USA}

\author{Gia-Wei Chern}
\affiliation{Department of Physics, University of Virginia, Charlottesville, VA 22904, USA}

\date{\today}

\begin{abstract}
We develop point-group equivariant neural networks for learning local mappings used in dynamical simulations of lattice systems. The same network acts on a finite neighborhood around each site, respecting lattice translations through shared parameters. The local configuration is decomposed into features that transform according to irreducible representations of the point group, and hidden features retain this organization throughout the network. Equivariant linear maps and symmetry-allowed tensor products then learn nonlinear spatial correlations while preserving the required transformation laws. This approach avoids constructing a fixed set of hand-crafted invariant descriptors, whose selection requires choices about which local correlations to retain. It also allows the complexity of the learned mapping to grow without enlarging its spatial cutoff. We demonstrate the construction by learning the forces governing adiabatic dynamics in the square-lattice Holstein model. Comparisons with exact diagonalization test the accuracy of the predicted forces and the evolution of spatial correlations. Force evaluation scales linearly with system size, allowing the trained model to simulate charge-density-wave ordering on large lattices. These simulations reproduce anomalously slow coarsening of the charge-density-wave domains, illustrating the use of the learned local mapping for collective dynamics.
\end{abstract}

\maketitle

\section{Introduction}
\label{sec:intro}

Machine-learning (ML) interatomic potentials have become a powerful tool in quantum chemistry and atomistic materials modeling. Trained on electronic-structure data, they can reproduce quantum-mechanical energies and forces with high accuracy while substantially reducing the cost of force evaluation, making ab initio molecular dynamics (AIMD) practical for larger systems and longer times~\cite{behler07,bartok10,li15,behler16,shapeev16,zhang18,deringer19,unke21}. Related ML force-field approaches have also been developed for lattice Hamiltonians, including electron-lattice models~\cite{cheng23a,Ghosh24,Fan2026GNN} and electron-spin systems such as the $s$-$d$ model~\cite{zhang21,Fan24,zhang23,cheng23b,Chern2026JMMM}. These approaches provide a route to studying collective dynamics in systems with classical fields coupled to electrons beyond the scales accessible to direct electronic simulations.

Locality provides the physical basis for scalable ML force fields. Motivated by the nearsightedness of electronic matter~\cite{Kohn1996,Prodan2005}, the energy or force associated with a site is approximated as a function of the configuration within a finite neighborhood. The Behler--Parrinello (BP) framework implements this idea by expressing the total energy as a sum of local contributions, $E=\sum_i\epsilon_i$, with each contribution predicted by a neural network from its local environment~\cite{behler07,behler16}. Forces follow by differentiating the learned total energy, although the same locality assumption also supports models that predict forces directly. Applying a shared local model throughout the system enables transfer between different system sizes and yields a computational cost proportional to the number of sites at fixed interaction range and model complexity. This local construction remains the basis of many modern force-field architectures, even as the representation of the environment and the neural network used to process it have evolved.

Within this local framework, symmetry determines how a model should respond to transformations of its input environment. In atomistic force fields, the relevant spatial symmetry is the Euclidean group $E(3)$, comprising continuous translations and rotations together with reflections. The energy must remain invariant under these transformations, while the forces must transform consistently with their vector character. For a lattice Hamiltonian, the underlying lattice restricts the spatial symmetry to discrete translations and a finite set of point-group operations that preserve the Hamiltonian. These symmetries likewise constrain a learned mapping on the lattice: transforming the input configuration must produce the corresponding transformation of its output.

A conventional approach to incorporating spatial symmetry is to represent local environments through invariant features. Atom-centered symmetry functions (ACSFs) established an influential route to this construction in neural-network potentials~\cite{behler07,behler11,smith17}. A broad range of approaches has since emerged, encompassing neighbor-density expansions, moment-tensor contractions, and learned local representations that encode many-body correlations~\cite{bartok10,bartok13,thompson15,shapeev16,schutt17,lubbers18,zhang18}. For systematic descriptor construction, the atomic cluster expansion (ACE) provides a general framework with a systematically improvable hierarchy of invariant basis functions organized by body order and spatial resolution~\cite{drautz19}.

For condensed-matter lattice models, group-theoretical descriptors incorporate the finite point group of the local environment~\cite{zhang22}. In the bispectrum approach~\cite{kondor2007,kondor2007skew,Kondor25}, the neighborhood configuration is decomposed into irreducible representations (IRs), whose coefficients are coupled through Clebsch--Gordan coefficients to form invariants. These invariants encode relative phases between symmetry components beyond their individual amplitudes. A reference-IR construction retains this information more compactly by comparing the components with reference representations derived from the same environment~\cite{zhang22,cheng23a}. Such descriptors have enabled practical local force-field models with exact point-group symmetry~\cite{cheng23a,Ghosh24,Fan24}, while requiring explicit choices about which invariant combinations to retain.

Equivariant neural networks (ENNs) extend this symmetry-based construction by learning representations of the local environment while preserving their transformation laws throughout the network~\cite{Cohen16,Kruger23,Kondor25}. Rather than reducing the input to a prescribed set of invariants, ENNs retain features that transform under nontrivial representations of the symmetry group and combine them through equivariant linear maps and symmetry-allowed nonlinear operations. The final output is then constructed to have the symmetry required by the predicted quantity. This approach has enabled accurate atomistic force fields that respect continuous rotational and Euclidean symmetries~\cite{Batatia22,batzner2022,musaelian2023}.

Implementations of ENNs differ in how they organize spatial information. Equivariant message-passing architectures exchange learned features between neighboring atoms, allowing successive layers to develop correlations over an expanding spatial neighborhood~\cite{Batatia22,batzner2022}. Strictly local architectures, such as Allegro, instead build nonlinear many-body correlations within a prescribed local environment, so that increasing network depth does not enlarge the spatial range of the local energy model~\cite{musaelian2023}. Equivariance is therefore compatible with both message passing and learning within a fixed spatial cutoff. The latter provides a natural starting point for combining the locality of the BP framework with learned representations adapted to discrete lattice symmetries.

In this work, we develop a point-group ENN for learning local mappings of lattice configurations by combining the BP construction with equivariant feature processing. The same network acts on finite neighborhoods centered on translationally equivalent lattice sites, ensuring translation equivariance, while its internal operations respect the point-group symmetry of the local environment. The neighborhood configuration is decomposed into irreducible representations, and hidden features retain this representation structure as they are processed through equivariant linear maps, symmetry-allowed tensor products, and nonlinear activations. The resulting local mapping can be used to predict quantities with the appropriate symmetry, including forces or the configuration after a prescribed time interval. Its spatial cutoff fixes the range of the input environment, while the complexity of the learned mapping can be increased without enlarging that range.

The formulation applies to different lattice geometries and spatial dimensions once the relevant point group and its action on the local environment are specified. Here we demonstrate it by learning the forces governing adiabatic dynamics in the square-lattice Holstein model. Benchmarks against exact diagonalization assess the accuracy of both the predicted forces and the resulting dynamics. The trained model then enables large-scale simulations of charge-density-wave ordering and its slow coarsening, demonstrating the practical use of the equivariant local mapping as a lattice force field.

The remainder of the paper is organized as follows. Section~\ref{sec:ENN} presents the general construction and its relation to descriptor-based and message-passing approaches. Section~\ref{sec:holstein} introduces the Holstein model, specializes the ENN to the square lattice, and benchmarks its force predictions and dynamics. Section~\ref{sec:coarsening} presents large-scale simulations of CDW coarsening, and Sec.~\ref{sec:discussion} concludes with a summary and outlook.

\section{Point-group equivariant mappings on lattices}
\label{sec:ENN}

In this section, we formulate a framework for local equivariant mappings with lattice symmetry built into its architecture. We first define the local environment and its transformation under the lattice point group, then construct an equivariant neural network using irreducible representations and symmetry-allowed couplings. We also discuss how this construction relates to invariant-descriptor and message-passing approaches. The formulation is presented for a general lattice geometry before being specialized to the square-lattice Holstein model in Sec.~\ref{sec:holstein}.

\subsection{Local environments and lattice symmetry}

The representation-based ENN construction presented below can be generalized to other discrete symmetry groups. Here we focus on learning local mappings that respect lattice translations and point-group symmetry. Such mappings can predict forces or directly approximate the evolution of a classical field over a prescribed time interval. We first present the general symmetry construction and subsequently demonstrate its application to force prediction in the semiclassical Holstein model.

Consider a lattice in $d$ spatial dimensions with sites at positions $\mathbf r_i$ and a real field $X_i$ at each site. We assume that point-group operations permute the sites while the on-site field transforms trivially. A representative example is the local distortion coordinate in the Holstein model~\cite{Holstein1959}, which describes a fully symmetric mode coupled to the electron density and can represent a local breathing distortion~\cite{Verges2002}. The same spatial transformation law applies to local electronic densities describing charge order in the $t$-$V$ model~\cite{MaTong2021,Yang2024CDW} and the Falicov--Kimball model~\cite{Chen2003FK}, as well as effective scalar order parameters near the finite-temperature Mott critical endpoint~\cite{Kotliar2000}. Thus, the field may represent either a microscopic classical coordinate or a collective electronic variable, while the symmetry constraints on the learned mapping have the same form.

For a cutoff radius $r_c$, let $\mathcal N_i(r_c)=\{j\mid |\mathbf r_j-\mathbf r_i|\le r_c\}$ denote the neighborhood of site $i$. We define its configuration as
\begin{equation}
    \mathcal C_i
    =\{(\mathbf r_j-\mathbf r_i,X_j)\mid j\in\mathcal N_i(r_c)\}.
    \label{eq:local-environment}
\end{equation}
Here $X_i$ is a classical scalar field on the lattice. Each field value in $\mathcal C_i$ is paired with its relative lattice position, so the spatial arrangement is part of the input. A learned local map takes the form
\begin{equation}
    Y_i=\mathcal M_{\bm\theta}(\mathcal C_i), \qquad Y_i\in V_{\mathrm{out}},
    \label{eq:local-map}
\end{equation}
where $\bm\theta$ denotes the trainable parameters and $V_{\mathrm{out}}$ specifies how the output transforms under symmetry. Depending on the application, $Y_i$ may be a scalar, a vector, or a collection of components forming another physical quantity. Applying the same map to all translationally equivalent sites enforces translation equivariance and permits transfer between system sizes. For a fixed cutoff and network architecture, the cost of evaluating the map at every site grows linearly with the number of sites. Lattices with inequivalent sites can be treated using separate local maps for the corresponding site types, with parameters shared among symmetry-equivalent environments.

Let $G$ be the finite point group of the reference environment about its central site. Its operations must preserve both the background lattice Hamiltonian and the selected neighborhood. For a lattice with inequivalent sites in its basis, $G$ is the appropriate site-symmetry group. Since the on-site field is a scalar, a point-group operation changes the positions of the field values without changing the values themselves:
\begin{equation}
    R\mathcal C_i
    =\{(\mathbf O(R)(\mathbf r_j-\mathbf r_i),X_j)
      \mid j\in\mathcal N_i(r_c)\},
    \qquad R\in G,
    \label{eq:group-action}
\end{equation}
where $\mathbf O(R)$ is an orthogonal matrix acting on physical space. We require the local map to satisfy
\begin{equation}
    \mathcal M_{\bm\theta}(R\mathcal C_i)
    =\rho_{\mathrm{out}}(R)\mathcal M_{\bm\theta}(\mathcal C_i),
    \qquad R\in G,
    \label{eq:output-equivariance}
\end{equation}
where $\rho_{\mathrm{out}}$ is the chosen representation on $V_{\mathrm{out}}$. The output may belong to a single irreducible representation (IR) or contain several IR components. In the latter case, the components can be assembled into a physical vector or tensor when their combined transformation law matches that quantity. For an invariant scalar output, $\rho_{\mathrm{out}}(R)=1$, and the local map is point-group invariant. The collection of site outputs nevertheless transforms equivariantly under lattice operations that move the central site.

The cutoff neighborhood should contain complete symmetry orbits so that the group action is well defined. The construction does not require a particular coordination number or spatial dimension: square, triangular, and cubic environments, for example, enter through their site geometry and point-group representations.

\begin{figure*}[t]
\centering
\includegraphics[width=1.99\columnwidth]{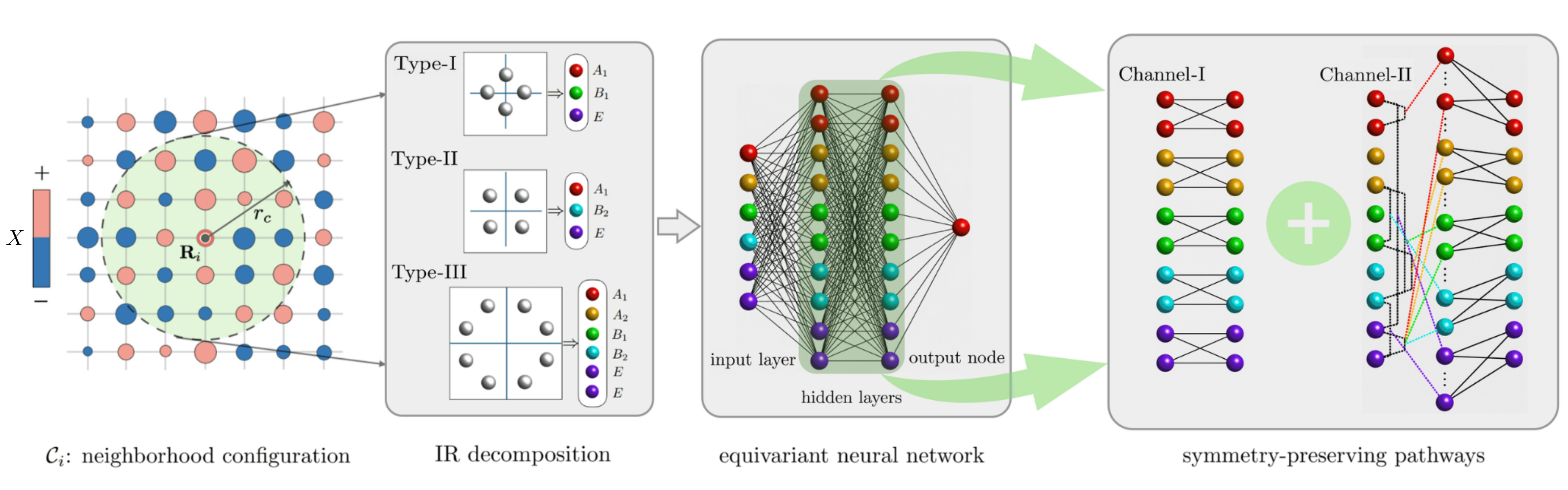}
\caption{Point-group equivariant neural-network framework for local lattice force fields, illustrated with a square-lattice environment. The scalar fields in a finite neighborhood $\mathcal C_i$ are decomposed into irreducible representations (IRs) of its point group, $D_4$ for the example shown. Hidden features retain their prescribed transformation laws, and the invariant sector supplies the scalar force prediction. Equivariant updates combine two pathways: linear mixing of features within an IR sector and tensor-product coupling followed by projection onto a target IR using Clebsch--Gordan coefficients. The resulting features are linearly mixed within each target sector and passed through equivariant nonlinear activations. The same construction applies to other lattice geometries with their corresponding point groups and representation data.}
\label{fig:ml-scheme}
\end{figure*}

\subsection{Symmetry-adapted representations of the neighborhood}

Because each field value in $\mathcal C_i$ is associated with a relative lattice position, the action of $G$ permutes the input values among those positions. This permutation action generally defines a reducible representation. A change of basis decomposes the local configuration into irreducible representation (IR) sectors~\cite{hamermesh_group_1989}. We group the coefficients belonging to each copy of an IR into a vector $\bm f^{(\Gamma,r)}\in V_\Gamma$. In an orthonormal real representation basis, its components are
\begin{equation}
    f_a^{(\Gamma,r)}
    =\sum_{j\in\mathcal N_i(r_c)}U_{aj}^{(\Gamma,r)}X_j,
    \qquad a=1,\ldots,d_\Gamma,
    \label{eq:IR_def}
\end{equation}
where $\Gamma$ labels the IR type, $d_\Gamma$ is its dimension, and $r$ distinguishes repeated copies. The coefficients $U_{aj}^{(\Gamma,r)}$ depend on the positions in the chosen neighborhood and the action of $G$ on those positions. Real representation blocks can be used for real input fields, combining conjugate complex representations when necessary. If all vectors $\bm f^{(\Gamma,r)}$ are retained, the decomposition is a change of coordinates on the finite input space and discards no information.

Point-group operations preserve the distance from the central site, so different radial shells do not mix under the group action. The decomposition can therefore be performed independently for each shell, or further separated into symmetry orbits when a shell contains several inequivalent sets of sites. A given IR may occur in multiple shells or orbits. The corresponding vectors $\bm f^{(\Gamma,r)}$ describe distinct spatial patterns with the same transformation law and can subsequently be coupled by the network.

We denote the features processed by the network as $\bm x^{(\Gamma,r)}$. At the input layer these are the vectors $\bm f^{(\Gamma,r)}$; in subsequent layers, $r$ labels learned feature channels belonging to the same IR. Each such feature transforms as
\begin{equation}
    \bm x^{(\Gamma,r)}\longrightarrow
    \bm x_R^{(\Gamma,r)}
    =\bm{\mathcal A}_\Gamma(R)\bm x^{(\Gamma,r)},
    \qquad R\in G,
    \label{eq:input-f}
\end{equation}
where $\bm{\mathcal A}_\Gamma(R)$ is an orthogonal representation matrix. Although each original $X_j$ is a scalar, its arrangement across the neighborhood gives rise to input vectors in nontrivial IR sectors. Maintaining these sectors in the hidden layers allows the network to process spatial patterns according to how they rotate or reflect, before forming an output with the required transformation law. Figure~\ref{fig:ml-scheme} illustrates the decomposition for the square lattice used below.

\subsection{Equivariant layers and output readout}

Each layer combines two equivariant pathways. Channel-I propagates existing features through linear mixing of copies of the same IR. Channel-II forms nonlinear products of features and projects them onto allowed target IRs. Specifically, if $\Gamma$ occurs in $\Gamma_1\otimes\Gamma_2$, a mixed feature can be constructed as
\begin{equation}
    \widetilde{x}^{(\Gamma,r)}_a
    =\sum_{bc}C^{(\Gamma,\mu;\Gamma_1,\Gamma_2)}_{abc}
    x_b^{(\Gamma_1,r_1)}x_c^{(\Gamma_2,r_2)},
    \label{eq:tensor-product}
\end{equation}
where $C$ contains the Clebsch--Gordan coefficients~\cite{Kondor25,hamermesh_group_1989}. The index $\mu$ distinguishes multiple occurrences of the target IR in a tensor-product decomposition when necessary; the output multiplicity label $r$ includes the chosen input channels and coupling. The two input IR types may be identical or different. Projection onto $\Gamma$ ensures that the product has a definite transformation law. Repeated layers can thus construct nonlinear correlations involving progressively more input features while remaining within the original neighborhood.

For a given target IR, the direct and mixed features form an expanded set, symbolically $\bm z=\bm x\oplus\widetilde{\bm x}$. An equivariant linear map then mixes copies of that IR,
\begin{equation}
    Y_a^{(\Gamma,r)}
    =\sum_{r'}W_{r,r'}^{(\Gamma)}z_a^{(\Gamma,r')}.
    \label{eq:ir-mixing}
\end{equation}
Applying the same weights to every component $a$ preserves the transformation law of the IR. The architecture specifies which channels and tensor-product couplings are retained; their multiplicities control the size and computational cost of the intermediate representation. This provides a way to vary nonlinear capacity without changing $r_c$.

We introduce a further nonlinearity by applying a scalar activation to the amplitude of each feature,
\begin{equation}
    \bm y^{(\Gamma,r)}
    =\sigma\!\left(\lVert\bm Y^{(\Gamma,r)}\rVert+b_r^{(\Gamma)}\right)
    \hat{\bm Y}^{(\Gamma,r)},
    \label{eq:activation}
\end{equation}
Here $\hat{\bm Y}^{(\Gamma,r)}=\bm Y^{(\Gamma,r)}/\lVert\bm Y^{(\Gamma,r)}\rVert$ is the unit vector along the feature, $b_r^{(\Gamma)}$ is a trainable scalar parameter, and $\sigma(\cdot)$ is the chosen activation function. Thus, the activation applies a nonlinear function to the feature's amplitude and multiplies the result by its direction. At zero norm, the activated feature is defined as zero. Because the representation matrices are orthogonal, the norm is invariant under $G$, while $\hat{\bm Y}^{(\Gamma,r)}$ transforms in the same IR as $\bm Y^{(\Gamma,r)}$. The activation therefore preserves equivariance.

The readout is determined by the desired output representation $\rho_{\mathrm{out}}$. It selects the corresponding IR sectors from the final layer and combines copies within each sector through equivariant linear maps to produce $Y_i\in V_{\mathrm{out}}$. If the output is expressed as a physical vector or tensor, a fixed change of basis can assemble these IR components into its usual spatial components. For an invariant scalar output, only the trivial IR contributes; on the square lattice considered below, this sector is denoted $A_1$. The readout therefore satisfies Eq.~\eqref{eq:output-equivariance}, while sharing the local network among translationally equivalent sites gives the corresponding translation equivariance.

\subsection{Relation to invariant descriptors and message passing}

The present construction shares the finite-neighborhood formulation of descriptor-based models, but differs in how it represents nonlinear correlations. For an invariant scalar output, a descriptor model first constructs a prescribed collection of invariant features $\bm g_i=\bm g(\mathcal C_i)$ and then learns
\begin{equation}
    \zeta_i=\mathrm{MLP}_{\bm\theta}(\bm g_i),
    \qquad \zeta_i\in\mathbb R.
    \label{eq:descriptor-map}
\end{equation}
The prediction thus depends on $\mathcal C_i$ only through $\bm g_i$. Invariant inputs and an ordinary scalar readout ensure point-group invariance, but cannot by themselves produce an output that transforms in a nontrivial IR.

The IR coefficients introduced above provide a systematic way to construct $\bm g_i$. In the bispectrum approach, three IR components are coupled to form an invariant~\cite{kondor2007,zhang22}:
\begin{equation}
    B_{r_1r_2r_3}^{(\Gamma_1\Gamma_2\Gamma_3)}
    =\sum_{abc}C_{abc}^{\Gamma_1\Gamma_2\Gamma_3}
    f_a^{(\Gamma_1,r_1)}f_b^{(\Gamma_2,r_2)}f_c^{(\Gamma_3,r_3)}.
    \label{eq:bispectrum-invariant}
\end{equation}
Here $C_{abc}^{\Gamma_1\Gamma_2\Gamma_3}$ denotes a Clebsch--Gordan coupling of $\Gamma_1\otimes\Gamma_2$ to $\Gamma_3$; an additional label is understood if the coupling has multiplicity. The quadratic invariant $I_{rs}^{(\Gamma)}=\sum_a f_a^{(\Gamma,r)}f_a^{(\Gamma,s)}$ is included by taking the third factor to be a fixed unit feature in the trivial IR. Its diagonal terms $I_{rr}^{(\Gamma)}=\lVert\bm f^{(\Gamma,r)}\rVert^2$ give the power spectrum. Bispectrum terms can additionally retain relative orientation or phase information between IR components. The reference-IR approach encodes such information with fewer explicitly constructed invariants by comparing a component with a reference that transforms in the same way and is derived from the local environment~\cite{zhang22,cheng23a}.

The choice of which invariants to include determines what the MLP can learn. It can form flexible nonlinear combinations of $\bm g_i$, but cannot recover distinctions between configurations that have identical descriptor vectors. Retaining more invariants may address this problem at the cost of a larger input representation, so a practical descriptor model requires choices about which correlations to construct in advance. By contrast, the ENN initially retains the full IR decomposition of $\mathcal C_i$ and learns symmetry-allowed combinations across its hidden layers. Its readout can also select the IR sectors required for a non-scalar output.

Message-passing graph neural networks (GNNs) provide another way to learn local representations without prescribing invariant descriptors~\cite{scarselli09,battaglia2018,xie18,Chen2019}. The lattice is represented as a graph, and each site updates its features using messages from neighboring sites. Applying the same update rule at translationally equivalent sites respects discrete translations; sharing weights among symmetry-equivalent edges and aggregating their messages symmetrically also respects the lattice point group. This strategy has been applied to a GNN force field for the Holstein model~\cite{Fan2026GNN}. The spatial range of a message-passing model is controlled by its propagation steps: on a nearest-neighbor graph, features after $L$ steps can depend on sites up to $L$ edges away. Adding propagation steps to increase expressive power therefore also enlarges the receptive field. Additional nonlinear processing within each step, such as MLPs before and after aggregation, can increase capacity while holding the number of propagation steps fixed~\cite{Fan2026GNN}. In the present ENN, the chosen neighborhood is available from the input layer, and additional layers construct nonlinear correlations within that same spatial cutoff.

A simple example illustrates why the organization of local information matters. Consider a direction-independent aggregation of the four nearest-neighbor values on a square lattice, $\sum_{j\in\mathcal N(i)}h(X_j)$, for any shared scalar function $h$. It gives the same result for $(X_N,X_E,X_S,X_W)=(1,1,0,0)$ and $(1,0,1,0)$, although the nonzero values occupy adjacent sites in the first configuration and opposite sites in the second. These configurations are not related by $D_4$. The symmetry-adapted component $X_N+X_S-X_E-X_W$ vanishes for the first configuration and equals $2$ for the second; its square is an invariant that preserves the distinction. This example concerns a single direction-independent aggregation, rather than a limitation of GNNs in general: further message passing or richer geometric features can distinguish additional patterns. The ENN makes the spatial symmetry sectors explicit from the outset and learns their nonlinear couplings at a prescribed range. Equivariant features can also be incorporated into message-passing architectures~\cite{Batatia22,batzner2022,musaelian2023}.

\section{Application to the square-lattice Holstein model}
\label{sec:holstein}

We now apply the general construction to adiabatic dynamics of the square-lattice Holstein model. The lattice fixes the relevant site-symmetry group to $D_4$, while the dynamics supplies a scalar local force to be learned. We first introduce the Hamiltonian and its charge-ordered state, then specialize the symmetry-adapted input and network readout and benchmark the resulting force field.

\subsection{Hamiltonian, point-group symmetry, and charge order}

\begin{figure}[t]
\centering
\includegraphics[width=0.99\columnwidth]{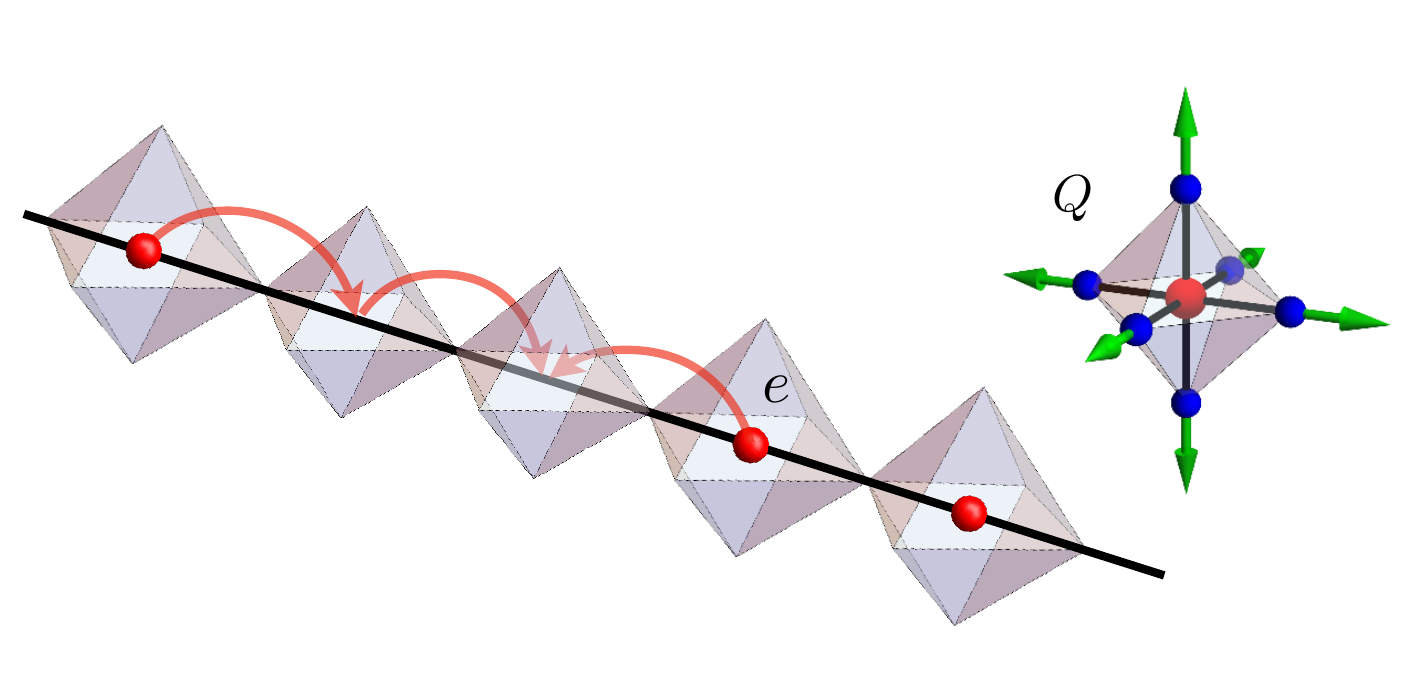}
\caption{Square-lattice Holstein model used to demonstrate the local point-group ENN force field. Itinerant electrons couple to scalar variables $Q_i$ describing local $A_1$-type structural distortions. Rotations and reflections permute these site variables, while their spatial arrangement supplies nontrivial $D_4$ representations to the ENN.}
\label{fig:holstein}
\end{figure}

The semiclassical Holstein model~\cite{Holstein1959} provides a standard setting for studying electron--phonon coupling, polaron formation, and charge ordering~\cite{noack91,zhang19,esterlis19,chen19,hohenadler19,bonca99,golez12,mishchenko14,scalettar89,costa18,bradley21,li2025}. It describes itinerant electrons coupled to local distortions $Q_i$, as illustrated in Fig.~\ref{fig:holstein}. On the square lattice, the Hamiltonian is
\begin{align}
    \hat{\mathcal H}
    &=\hat{\mathcal H}_{\rm el}+\sum_i\frac{P_i^2}{2m}+V_{\rm lat},\notag\\
    \hat{\mathcal H}_{\rm el}
    &=-t_{\rm nn}\sum_{\langle ij\rangle}\hat c_i^\dagger\hat c_j
      -g\sum_i Q_i\hat n_i,\notag\\
    V_{\rm lat}
    &=\frac{k}{2}\sum_i Q_i^2
      +\kappa\sum_{\langle ij\rangle}Q_iQ_j.
    \label{eq:H_holstein}
\end{align}
Here $\hat c_i^\dagger$ ($\hat c_i$) creates (annihilates) an electron at site $i$, and $\hat n_i=\hat c_i^\dagger\hat c_i$ is the local electron number operator. The hopping sum includes the Hermitian-conjugate contribution. The coordinate $Q_i$ represents the amplitude of a local distortion, such as a breathing mode of an atomic cluster, and $P_i$ is its conjugate momentum. The parameters $t_{\rm nn}$ and $g$ set the nearest-neighbor hopping and electron--lattice coupling. The classical lattice sector has mass $m$, harmonic stiffness $k$, and an antiferrodistortive coupling $\kappa$ between neighboring modes. Thus, the generic scalar field $X_i$ introduced in Sec.~\ref{sec:ENN} is identified with $Q_i$ in this application.

The local point group is $D_4$, generated by a fourfold rotation and a reflection. Each on-site displacement transforms in its trivial $A_1$ representation, as does the force conjugate to it. A finite neighborhood nevertheless contains spatial patterns in nontrivial IR sectors, which the ENN retains before the final $A_1$ readout. The Holstein model thus isolates the role of point-group equivariance in learning the geometry of scalar lattice environments. The required symmetry acts on the arrangement of the neighboring distortions, without an additional transformation of internal on-site components.

At half filling, the square-lattice model supports a finite-temperature transition into a CDW phase~\cite{noack91,zhang19,esterlis19,chen19,hohenadler19}. Its two sublattices have densities $n_{A/B}=(1\pm\delta)/2$, accompanied by staggered distortions $Q_{A/B}=\bar Q\pm\mathcal Q$, where $\bar Q$ denotes the uniform equilibrium displacement. Interchanging the two sublattices reverses the sign of the staggered order parameter. The associated discrete symmetry breaking gives an Ising transition in the parameter regimes established by previous studies~\cite{noack91,zhang19,esterlis19}. Following a thermal quench, domains of the two checkerboard patterns form and evolve through domain-wall motion. Their slow coarsening provides a demanding dynamical test of local force predictions, since errors can accumulate over the long times needed to resolve domain growth.

\subsection{Adiabatic dynamics and electronic forces}

In the adiabatic approximation, the electrons equilibrate rapidly relative to the evolution of the classical distortions. For a fixed configuration, let $\mathcal V(\{Q_i\})$ be the effective potential governing those distortions, including $V_{\rm lat}$ and the electronic free energy; at zero electronic temperature the latter reduces to the electronic ground-state energy. Coupling the classical fields to a thermal reservoir gives the Langevin equation
\begin{equation}
    m\frac{d^2Q_i}{dt^2}
    =\mathcal F_i-\gamma\frac{dQ_i}{dt}+\eta_i(t),
    \qquad \mathcal F_i=-\frac{\partial\mathcal V}{\partial Q_i},
    \label{eq:langevin}
\end{equation}
where $\gamma$ is the damping coefficient. The Gaussian noise has zero mean and satisfies $\langle\eta_i(t)\eta_j(t')\rangle=2\gamma k_BT\delta_{ij}\delta(t-t')$, with $T$ the reservoir temperature.

The electronic contribution to the force is the equilibrium expectation value of the derivative of $\hat{\mathcal H}_{\rm el}$, as follows from the Hellmann--Feynman relation and its thermal counterpart~\cite{hellmann1937,feynman1939}. The resulting force is
\begin{equation}
    \mathcal F_i=-kQ_i-\kappa\sum_{j\in\mathcal N(i)}Q_j
      +g\langle\hat n_i\rangle,
    \label{eq:force-Holstein}
\end{equation}
where $\mathcal N(i)$ denotes the nearest neighbors of site $i$. The first two terms are explicit elastic forces. The final term requires the electronic density in the instantaneous distortion configuration. In the reference simulations, this density is obtained by exact diagonalization of the tight-binding Hamiltonian with on-site potentials $v_i^{\rm eff}=-gQ_i$. Repeating this calculation throughout the dynamics is the principal computational bottleneck.

For this force-field application, we choose the scalar output of the general map in Eq.~\eqref{eq:local-map} to represent the local force. With $X_i=Q_i$ and $V_{\mathrm{out}}=\mathbb R$, the ENN prediction is
\begin{equation}
    Y_i=\mathcal F_i^{\rm ENN}=\mathcal M_{\bm\theta}(\mathcal C_i),
    \label{eq:force-ML-func}
\end{equation}
where the training target is the reference force $\mathcal F_i$ in Eq.~\eqref{eq:force-Holstein}. The cutoff fixes the spatial information supplied to each prediction, while the $D_4$ decomposition and equivariant layers make the predicted force invariant under point-group operations about site $i$. The implementation predicts the force directly; differentiating a learned invariant energy would instead give a conservative force field.

One could also learn a finite-time update of the dynamical state, mapping local distortions and velocities to their values at a later time. Such a model might advance the simulation in larger time steps, but would have to learn the time-step-dependent evolution and account for the stochastic reservoir. By learning the force, we retain the Langevin equation, including its explicit damping and noise, and replace the repeated electronic calculation needed to evaluate its force term. We next describe the ENN implementation and benchmark its predicted forces and the resulting Langevin dynamics.

\subsection{ENN force-field implementation and benchmarks}
\label{sec:holstein-benchmarks}

To implement the local force map in Eq.~\eqref{eq:force-ML-func}, we use an environment $\mathcal C_i$ consisting of 45 sites within a cutoff radius $r_c=3.61$ lattice constants. The site distortions in this neighborhood are decomposed into IRs of the square-lattice point group $D_4$ before being processed by the ENN.

As discussed in Sec.~\ref{sec:ENN}, the local neighborhood decomposes naturally into radial-shell blocks, allowing the IR decomposition to be carried out independently for each block. The neighborhood-block decomposition and the conventions for its basis functions are given in Appendix~\ref{app:ir-decomposition}.  As a representative example, the four nearest neighbors $\{Q_a, Q_b, Q_c, Q_d \}$ of a given site form a type-I block whose decomposition under $D_4$ is given by $4 = 1A_1 \oplus 1B_1 \oplus 1E$. The corresponding unnormalized symmetry-adapted basis functions are $f^{(A_1)} = Q_a + Q_b + Q_c + Q_d$, $f^{(B_1)} = Q_a - Q_b + Q_c - Q_d$, and $\bm f^{(E)} = \left(Q_a - Q_c, Q_d - Q_b \right)$. The character table is given in Table~\ref{table:D4-charac}.

\begin{table}
\caption{Architecture of the ENN for the 2D Holstein model. Entries give channel multiplicities in the indicated IR order; each $E$ channel has two components. Hidden layers use ReLU for the amplitude activation $\sigma$ in Eq.~\eqref{eq:activation}.}
\label{tab:enn-arch}
\small
\begin{ruledtabular}
\begin{tabular}{l c c  c}
Layer  &\,\,& \makecell[c]{IR Channels \\ $A_1 \oplus A_2 \oplus B_1 \oplus B_2 \oplus E$} & Nonlinearity \\
\hline
Input & \,\,& $9 \oplus 3 \oplus 6 \oplus 5 \oplus 11$ & -- \\
Expanded 1 & \,\,&  $128 \oplus 115 \oplus 130 \oplus 123 \oplus 264$ & -- \\
Hidden 1 & \,\,&  $32 \oplus 32 \oplus 32 \oplus 32 \oplus 32$ & ReLU \\
Expanded 2 & \,\,& $2512 \oplus 2576 \oplus 2576 \oplus 2576 \oplus 4128$ & -- \\
Hidden 2 & \,\, & $8 \oplus 8 \oplus 8 \oplus 8 \oplus 8$ & ReLU \\
Expanded 3 &\,\, & $148 \oplus 164 \oplus 164 \oplus 164 \oplus 264$ & -- \\
Hidden 3 &\,\,& $4 \oplus 4 \oplus 4 \oplus 4 \oplus 4$ & ReLU \\
Expanded 4 &\,\,& $34 \oplus 42 \oplus 42 \oplus 42 \oplus 68$ & -- \\
Readout & \,\,&  $1 \oplus 0 \oplus 0 \oplus 0 \oplus 0$ & Linear
\end{tabular}
\end{ruledtabular}
\end{table}

Aggregating contributions from all radial shells, the classical fields in $\mathcal{C}_i$ decompose as: $45 = 9A_1 \oplus 3A_2 \oplus 6B_1 \oplus 5B_2 \oplus 11E$. With this input dimension, the detailed architecture of the ENN employed in this work is summarized in Table~\ref{tab:enn-arch}.   The network is implemented in PyTorch, with the features at each stage organized according to the \(A_1 \oplus A_2 \oplus B_1 \oplus B_2 \oplus E\) IR sectors of the \(D_4\) point group. Implementation details are given in Appendix~\ref{app:enn-setup}.

The symmetry-expanded feature set at each stage combines (i) direct features that retain their IR type and (ii) symmetry-allowed inter-IR combinations generated through the tensor-product channel, as described in Sec.~\ref{sec:ENN}. These mixed features are constructed according to the \(D_4\) group multiplication rules using the corresponding Clebsch--Gordan coefficients; the direct-product rules and Clebsch--Gordan coefficients are given in Appendix~\ref{app:cg-d4}. Features belonging to the same IR sector are then propagated through fully connected transformations, followed by the equivariance-preserving nonlinear activation defined in Eq.~(\ref{eq:activation}) for the hidden layers. The final readout retains only the \(A_1\) sector, corresponding to the scalar Holstein force, with no activation function applied at the output. This construction preserves the \(D_4\) point-group symmetry of the force prediction while retaining the expressive power provided by symmetry-allowed inter-IR coupling.

\begin{figure}[t]
\centering
\includegraphics[width=0.99\columnwidth]{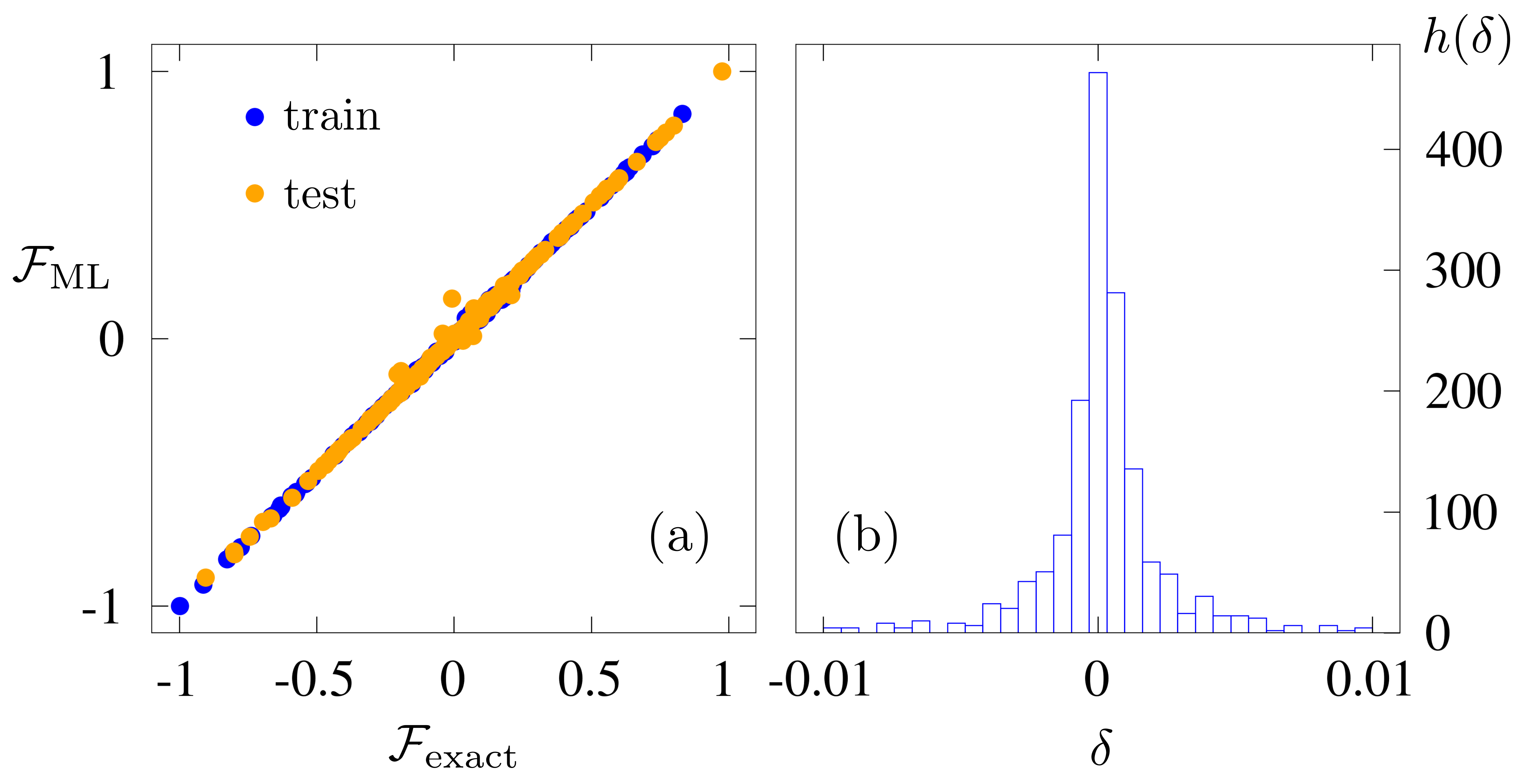}
\caption{(a) Scatter plot comparing the forces predicted by the ENN, $F_{\mathrm{ML}}$, with the exact forces $F_{\mathrm{exact}}$ obtained from exact diagonalization. Blue and orange symbols denote training and test configurations, respectively. (b) Histogram of the force prediction error $\delta = F_{\mathrm{ML}} - F_{\mathrm{exact}}$ for the test set, showing a narrow, approximately symmetric distribution centered around zero.}
    \label{fig:f-benchmark}
\end{figure}

The trainable parameters \(\bm \theta = \{ W^{(\Gamma)}_{r,r'}, b^{(\Gamma)}_{r{\phantom{'}}} \}\) consist of the weight matrices that map the symmetry-expanded features to the subsequent hidden or output layers, together with the corresponding bias terms. The resulting ENN contains a total of 143,135 trainable parameters. Since the ENN is designed to directly predict the local forces, the optimal parameters are obtained by minimizing the squared-error loss function
\begin{equation}
	\mathcal{L} = \sum_i \left| \mathcal{F}^{\rm ML}_i - \mathcal{F}^{\rm exact}_i \right|^2,
    \label{eq:force-loss}
\end{equation}
where $\mathcal{F}^{\rm exact}_i$ are the forces computed from exact diagonalization and the sum runs over the site predictions in a training batch. Dividing by the number of predictions gives the corresponding mean-squared error. The optimization is carried out using the Adam algorithm with a learning rate of 0.001. The training dataset is constructed from exact diagonalization (ED) solutions of a mixture of random lattice configurations and quasi-ordered CDW states on a $40 \times 40$ lattice. We employ a dimensionless electron-lattice coupling $\lambda = g^2/(kW) = 1.5$, where $W = 8 t_{\rm nn}$ denotes the electronic bandwidth. To ensure robust sampling of configurations relevant to nonequilibrium relaxation, intermediate states obtained from thermal quench simulations are included in the training set, yielding a total of 600 configurations, 200 of which are reserved for testing. 

\begin{figure}[t]
\centering
\includegraphics[width=0.9\columnwidth]{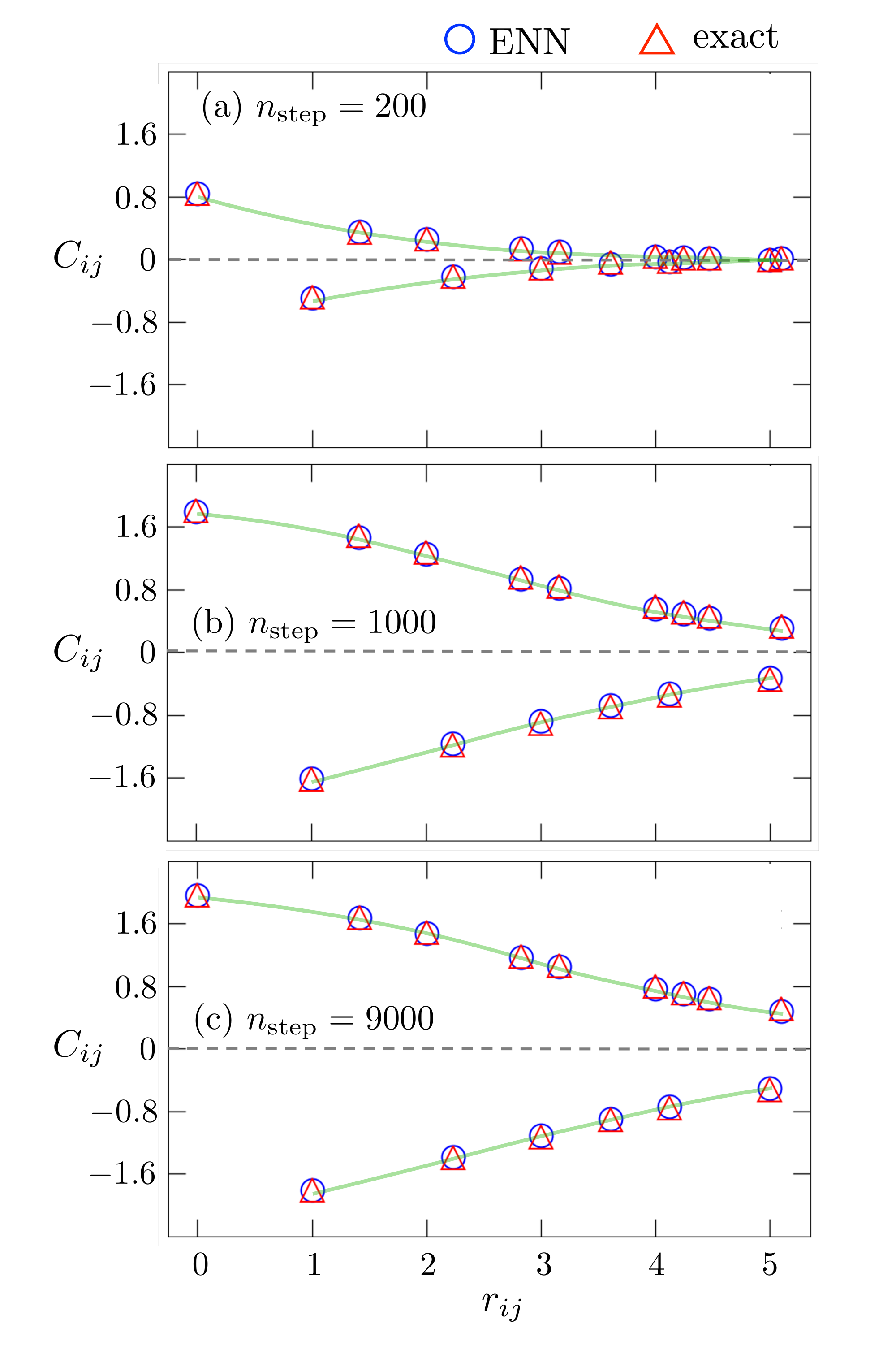}
\caption{Equal-time lattice correlation functions $C_{ij} = \langle Q_i Q_j \rangle$ as a function of the site separation $r_{ij}$ following a thermal quench to $T = 0.1$, shown at three representative times: (a) $n_{\rm step}=200$, (b) $n_{\rm step}=1000$, and (c) $n_{\rm step}=9000$. Blue circles denote results obtained from Langevin dynamics using ENN-predicted forces, while red triangles correspond to exact diagonalization (ED)–based Langevin simulations. Each data point is averaged over 30 independent runs to reduce statistical fluctuations. }
    \label{fig:corr-benchmark}
\end{figure}

Figure~\ref{fig:f-benchmark}(a) compares the ENN-predicted and exact forces across the sampled force range. The 400 training configurations supply $400\times40\times40$ local environments. These environments provide many site-resolved training targets, although overlapping neighborhoods and spatial correlations mean that they are not statistically independent samples. The split into training and test configurations allows force accuracy to be assessed on configurations excluded from training. The test-error distribution, $\delta=\mathcal F^{\rm ML}-\mathcal F^{\rm exact}$, is narrow and approximately centered at zero, with standard deviation $\sigma_\delta=0.0084$ [Fig.~\ref{fig:f-benchmark}(b)]. Together with the dynamical benchmarks below, these results assess the practical accuracy of the symmetry-constrained local model.

To assess whether the trained ENN can faithfully reproduce nonequilibrium dynamics beyond static force benchmarks, we incorporate the ML force model into Langevin dynamics simulations of the Holstein model and compare the results directly with ED-based Langevin simulations. Starting from an initially random configuration, the system is quenched at time $t=0$ to a low temperature $T = 0.1$. We monitor the time evolution of the equal-time lattice correlation function $C_{ij} = \langle Q_i Q_j \rangle$ at various stages following the quench. To reduce statistical fluctuations associated with finite lattice sizes, all correlation functions are obtained by averaging over 30 independent simulation runs.

As shown in Fig.~\ref{fig:corr-benchmark}, the correlation functions exhibit a characteristic short-period oscillation, reflecting the staggered lattice distortions $Q_i \sim (-1)^{x_i + y_i}$ associated with the checkerboard CDW order, superimposed on a slowly varying envelope that encodes the evolving correlation length. Short-range CDW correlations emerge rapidly following the quench, already becoming apparent at $n_{\rm step} = 200$. In contrast, the subsequent development of longer-range order proceeds much more slowly. Importantly, at all examined times, the correlation functions obtained from ML-driven Langevin dynamics are in excellent agreement with their ED counterparts, providing strong evidence that the ENN not only delivers accurate local force predictions but also reliably captures the emergent nonequilibrium dynamics of the Holstein model.

\section{Large-scale simulations of CDW coarsening}
\label{sec:coarsening}

\begin{figure*}[t]
\centering
\includegraphics[width=1.99\columnwidth]{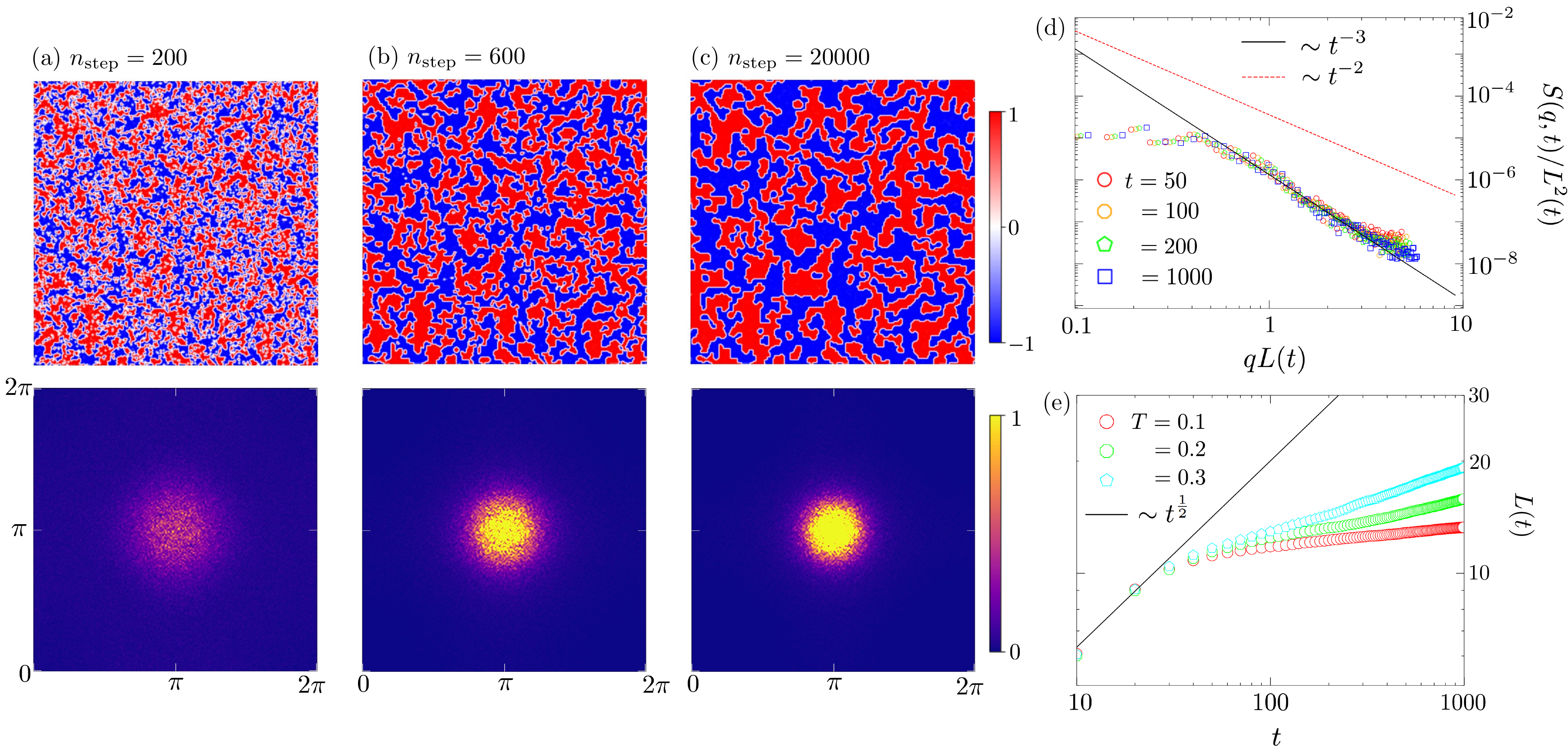}
\caption{Coarsening dynamics of the CDW phase in the Holstein model from ENN-based Langevin dynamics simulations.
(a)–(c) Real-space snapshots of the CDW order-parameter field (top) and the corresponding ensemble-averaged structure factors (bottom) at representative times following a thermal quench. (d) Scaled structure factor $S(q,t)/L^2(t)$ versus $qL(t)$ at different times, showing approximate data collapse over the displayed time window.  (e) Characteristic length scale $L(t)$ extracted from the structure factor as a function of time for three temperatures, showing slow growth with fitted apparent exponents smaller than the Allen--Cahn value $1/2$. }
    \label{fig:coarsening}
\end{figure*}

We use the trained ENN force field to follow charge-density-wave (CDW) ordering on lattices of up to $200\times200$ sites. Starting from disordered configurations with random local distortions, we quench the system to a temperature below the CDW transition at $t=0$ and evolve the subsequent Langevin dynamics using the same microscopic parameters as in the training data. These sizes and observation times allow us to examine domain growth well beyond the reach of simulations that repeatedly diagonalize the electronic Hamiltonian.

The electronic density along each ENN-driven trajectory is reconstructed from the predicted total force and the instantaneous lattice distortions using Eq.~\eqref{eq:force-Holstein}:
\begin{equation}
    n_i^{\rm ENN}
    =\frac{1}{g}\left(
    \mathcal F_i^{\rm ENN}+kQ_i
    +\kappa\sum_{j\in\mathcal N(i)}Q_j
    \right).
\end{equation}
This relation provides the evolving charge distribution without an additional electronic calculation. Denoting the reconstructed density simply by $n_i$, we define a local CDW order parameter:
\begin{equation}
    \phi_i=\left(n_i-\frac{1}{4}\sum_{j\in\mathcal N(i)}n_j\right)
    e^{i\mathbf Q\cdot\mathbf r_i}.
\end{equation}
where $\mathbf Q=(\pi,\pi)$ is the checkerboard ordering wave vector. The density difference measures the contrast between a site and its nearest neighbors, while the staggered phase factor assigns the same sign to sites belonging to a given CDW domain. The two symmetry-related checkerboard patterns have opposite signs of $\phi_i$.

Figure~\ref{fig:coarsening}(a--c) shows representative snapshots after the quench. Red and blue indicate regions of opposite CDW order, while the pale regions trace domain walls and areas of weak local order. Well-developed checkerboard order appears locally before the domains have grown large, leaving a mosaic of the two symmetry-related patterns. Over longer times, domain walls move and smaller domains disappear, producing the progressively coarser texture visible in the snapshots.

We quantify the evolving pattern through the CDW structure factor $S(\mathbf k,t)=\langle|\tilde n(\mathbf k,t)|^2\rangle$, where $\tilde n$ is the Fourier transform of the electron density.  The lower panels of Fig.~\ref{fig:coarsening}(a--c) show a peak around $\mathbf Q$ that grows in intensity and narrows as the domains expand. We define a characteristic length from the inverse first moment of this peak,
\begin{equation}
    L^{-1}(t)=\frac{\sum_{\mathbf k}S(\mathbf k,t)|\mathbf k-\mathbf Q|}
                       {\sum_{\mathbf k}S(\mathbf k,t)}.
\end{equation}
This measure tracks the typical domain size and the extent of CDW correlations.

Rescaling the structure factor by $L(t)$ produces an approximate collapse of the curves at different times, as shown in Fig.~\ref{fig:coarsening}(d):
\begin{equation}
    \frac{S(q,t)}{L^2(t)}=\mathcal G\bigl(qL(t)\bigr),
    \label{eq:sqt}
\end{equation}
where $q=|\mathbf k-\mathbf Q|$. Such dynamical scaling indicates that the domain morphology is approximately organized by a single growing length~\cite{Bray1994,Onuki2002,Puri2009}; it does not, by itself, require $L(t)$ to follow a single power law. At intermediate wave vectors, the scaled structure factor is consistent with the $q^{-3}$ Porod tail expected for sharp interfaces in two dimensions. The softer decay at larger wave vectors lies outside this sharp-interface scaling regime, where finite wall width, lattice-scale structure, and thermal fluctuations can affect the spectrum; it does not by itself identify a distinct coarsening mechanism.

Figure~\ref{fig:coarsening}(e) shows that $L(t)$ grows slowly over the simulated time windows. Fits to $L(t)\sim t^\alpha$ give apparent exponents $\alpha=0.059$, $0.115$, and $0.155$ for quenches to $T=0.1$, $0.2$, and $0.3$, respectively. All are well below the Allen--Cahn value $\alpha=1/2$ expected for ordinary curvature-driven coarsening of a nonconserved Ising-like order parameter~\cite{Allen1972,Bray1994,Onuki2002,Puri2009}. Slow, temperature-dependent growth was also observed in earlier force-field simulations of the semiclassical Holstein model~\cite{cheng23a}. Because these exponents are fitted over finite time intervals, we regard them as effective slopes rather than established asymptotic growth exponents.

The small apparent exponents suggest that domain-wall motion is subject to constraints beyond the simplest Allen--Cahn description. In that description, curvature drives an interface with an approximately constant mobility. A lower but constant mobility would reduce the growth prefactor without changing the asymptotic exponent $1/2$. The observed slopes therefore motivate examining how the available relaxation pathways and effective wall mobility evolve with the CDW texture, rather than attributing the slow growth solely to a uniformly reduced rate.

In the Holstein model, advancing a domain wall requires a coordinated change in the lattice distortions and the associated electronic density. Although the staggered CDW order parameter is not conserved, the total electronic charge is fixed, and the charge redistribution accompanying an interface displacement can alter the energetic cost of neighboring distortions. Importantly, the present adiabatic dynamics equilibrates the electrons at each lattice configuration: it does not impose a separate carrier-hopping time scale. Any hindrance associated with carrier rearrangement must therefore enter through the electronic contribution to the effective potential and the resulting collective lattice-relaxation pathways. Cooperative kink hopping has been identified as a source of slow coarsening in a one-dimensional quasiclassical Holstein model~\cite{Jang2026Kinks}. This provides a motivation to investigate correlated rearrangements here, without establishing the same mechanism for the present two-dimensional system.

The increase of the fitted exponent with temperature is compatible with thermal activation facilitating such collective rearrangements. The current data, however, do not distinguish a temperature-dependent asymptotic power law from a more complicated slow-growth law over a finite observation window. Resolving this issue requires examining individual domain-wall events and the accompanying equilibrium charge distributions, together with the dependence of the fitted slopes on observation time and system size. The ENN supplies an efficient force model for these investigations while retaining the microscopic electron--lattice coupling in its reference data.

\section{Conclusion and Outlook}
\label{sec:discussion}

We have developed a local equivariant neural network (ENN) for lattice systems that combines a finite input neighborhood with explicit point-group symmetry. The same learned map is applied at symmetry-equivalent sites, ensuring consistency under lattice translations. Within each neighborhood, the input is decomposed into irreducible representations (IRs), and equivariant linear maps and tensor products build nonlinear features while preserving their transformation laws. The output can be chosen to transform according to the physical quantity being predicted. In this construction, the cutoff sets the spatial range, while the network architecture controls the complexity of the correlations learned within that range.

As a demonstration, we use a scalar force readout for the semiclassical square-lattice Holstein model. The trained ENN reproduces forces from exact diagonalization (ED) and reproduces the evolution of spatial correlations in ED-based Langevin simulations. Its linear scaling with system size then permits simulations of charge-density-wave (CDW) ordering on lattices far larger than those accessible to repeated diagonalization. These simulations reproduce the slow, temperature-dependent coarsening reported in earlier work~\cite{cheng23a}. The coarsening results illustrate the practical value of an accurate local force model for collective dynamics over large spatial and temporal scales, without assigning the fitted effective exponents an asymptotic interpretation.

The ENN shares its finite-neighborhood starting point with descriptor-based models~\cite{zhang22,cheng23a}. In those models, point-group invariants, such as bispectrum components or features constructed with reference IRs, are selected before a conventional network processes the environment. This can be an effective and economical strategy, but the selection determines which distinctions between local configurations remain available to the network. Our construction instead retains the symmetry-adapted input components and learns their nonlinear combinations within equivariant layers, with the readout selecting the required output symmetry. It thus moves much of the representation design from a prescribed descriptor set into the trainable architecture. The resulting flexibility concerns how correlations are represented and learned; a complete invariant descriptor can also retain all information needed for an invariant prediction.

Message-passing graph neural networks (GNNs) offer another natural treatment of lattice symmetry, as demonstrated for Holstein dynamics in Ref.~\cite{Fan2026GNN}. Shared updates accommodate translations, while symmetry-consistent edge operations and aggregation account for the point group. Message passing also provides a flexible way to assemble information from neighboring sites. Its propagation depth typically determines how far that information can travel, although nonlinear processing within each step can increase expressivity without adding propagation steps. The present ENN takes a different route: the entire selected neighborhood is available at the input, and subsequent layers refine its IR-resolved correlations without extending the cutoff. These approaches offer complementary ways to control spatial range and representational capacity.

An important extension concerns lattice models with multicomponent on-site variables, such as Jahn--Teller distortions, orbital degrees of freedom, and spins coupled to the lattice through spin--orbit interactions. In these settings, a point-group operation acts both on the positions in a neighborhood and on the components attached to each site. The input representation must account for this combined action, and the allowed feature couplings and output representations must be constructed accordingly. The same representation-based framework provides a starting point, but constructing a consistent basis and coupling scheme for these joint spatial and on-site transformations, and controlling the resulting channel growth, require further development and validation. We leave these extensions to future work.

A second direction is to combine IR-resolved equivariant features with message passing. Symmetry-consistent messages could communicate information between neighborhoods, while equivariant mixing and tensor products construct nonlinear correlations within each site representation. The IR mixing and tensor-product operations developed here could serve as its local processing blocks, provided that the message construction respects the combined transformation laws at the connected sites. More generally, testing how predictions and long-time dynamics depend on the cutoff, retained IR channels, and training regime would help establish where a fixed-neighborhood model is sufficient and where a broader architecture is needed.

\begin{acknowledgments}
This work was supported by the US Department of Energy Basic Energy Sciences under Contract No. DE-SC0020330. G.W.C. thanks Qimin Yan for insightful discussions on ENN for crystalline systems. The authors acknowledge Research Computing at The University of Virginia for providing computational resources and technical support that have contributed to the results reported within this publication. 
\end{acknowledgments}

\appendix
\section*{Appendix}

\section{IR decomposition for the $D_4$ point group}
\label{app:ir-decomposition}

For the Holstein implementation, the general ENN is specialized to the eight-element symmetry group of the square. We denote this group by $D_4$ and use its planar rotation-and-reflection realization, conventionally denoted $C_{4v}$. Its five IRs are summarized in Table~\ref{table:D4-charac}.

The set of local lattice distortions $\{Q_j\}$ within a neighborhood
$\mathcal{C}_i$ surrounding a central site $i$ furnishes, in general,
a high-dimensional representation of the $D_4$ point group. This
neighborhood representation may be decomposed into a direct sum of
IRs using standard group-theoretical techniques. Owing to the
discrete lattice geometry, this decomposition is considerably
simplified. In particular, because the distance between a neighboring
site $j$ and the central site $i$ is invariant under all $D_4$
symmetry operations, the neighborhood representation is naturally
block diagonal in radial shells. A shell can contain several symmetry orbits; the formulas below apply separately to each orbit.

\begin{table}[b]
\begin{ruledtabular}
\begin{tabular}{c c c c c c l}
IR & $e$ & $2C_4$ & $C_2$ & $2\sigma_v$ & $2\sigma_d$ & Basis examples \\
\hline
$A_1$ & 1 & 1 & 1 & 1 & 1 & $1$, $x^2+y^2$ \\
$A_2$ & 1 & 1 & 1 & $-1$ & $-1$ & $xy(x^2-y^2)$ \\
$B_1$ & 1 & $-1$ & 1 & 1 & $-1$ & $x^2-y^2$ \\
$B_2$ & 1 & $-1$ & 1 & $-1$ & 1 & $xy$ \\
$E$ & 2 & 0 & $-2$ & 0 & 0 & $(x,y)$ \\
\end{tabular}
\end{ruledtabular}
\caption{Character table for the planar $D_4$ action. Here $e$ is the identity, $\sigma_v$ denotes reflection about a coordinate axis, and $\sigma_d$ denotes reflection about a diagonal.}
\label{table:D4-charac}
\end{table}

For the square lattice, these invariant blocks fall into three
distinct noncentral orbit types. Below, the letters denote field values at the specified relative positions. The same Cartesian convention is used for every $E$ doublet, so equivalent copies can be mixed directly.

For a type-I block, the four-dimensional representation decomposes as
\[
4=A_1\oplus B_1\oplus E .
\]
For sites $(p,0),(0,-p),(-p,0),(0,p)$ with $p>0$, the corresponding unnormalized basis functions are
\begin{eqnarray*}
f^{(A_1)} &=& a+b+c+d,\\
f^{(B_1)} &=& a-b+c-d,\\
\bm f^{(E)} &=& (a-c,\,-b+d).
\end{eqnarray*}

For the type-II block, the representation instead decomposes as
\[
4=A_1\oplus B_2\oplus E ,
\]
for sites $(p,p),(p,-p),(-p,-p),(-p,p)$. Its unnormalized basis functions are
\begin{eqnarray*}
f^{(A_1)} &=& a+b+c+d,\\
f^{(B_2)} &=& a-b+c-d,\\
\bm f^{(E)} &=& (a+b-c-d,\;a-b-c+d).
\end{eqnarray*}

The third class, denoted as type-III, corresponds to an eight-site
orbit with generic coordinates $p>q>0$ and furnishes an eight-dimensional representation of
the $D_4$ point group. This representation decomposes as
\[
8=A_1\oplus A_2\oplus B_1\oplus B_2\oplus E\oplus E ,
\]
with $(a,b,c,d,e,f,g,h)$ assigned respectively to $(p,q),\allowbreak(p,-q),\allowbreak(q,-p),\allowbreak(-q,-p),\allowbreak(-p,-q),\allowbreak(-p,q),\allowbreak(-q,p),\allowbreak(q,p)$. The unnormalized combinations are
\begin{eqnarray*}
f^{(A_1)} &=& a+b+c+d+e+f+g+h,\\
f^{(A_2)} &=& a-b+c-d+e-f+g-h,\\
f^{(B_1)} &=& a+b-c-d+e+f-g-h,\\
f^{(B_2)} &=& a-b-c+d+e-f-g+h,\\
\bm f^{(E,1)} &=& (a+b-e-f,\,-c-d+g+h),\\
\bm f^{(E,2)} &=& (c-d-g+h,\;a-b-e+f).
\end{eqnarray*}

To obtain the orthonormal basis of Eq.~\eqref{eq:IR_def}, divide the type-I singlets by $2$ and its doublet by $\sqrt{2}$; divide all type-II features by $2$; and divide the type-III singlets by $\sqrt{8}$ and its doublets by $2$. These fixed normalizations do not change their symmetry types.

The central site forms a one-dimensional block transforming
according to the trivial $A_1$ representation. For the cutoff
$r_c=3.61$ used in this work, the neighborhood consists of
this central $A_1$ singlet together with three type-I orbits, two type-II orbits, and three type-III orbits. Combining these
contributions gives
\[
45=9A_1\oplus3A_2\oplus6B_1\oplus5B_2\oplus11E.
\]

The resulting basis supplies the input features of Eq.~\eqref{eq:IR_def}.

\section{Clebsch--Gordan coefficients for the $D_4$ point group}
\label{app:cg-d4}

Clebsch--Gordan (CG) coefficients project products of input features onto definite IRs. For $D_4$, products of one-dimensional IRs follow their character multiplication; products involving $E$ require the component couplings given below.

Tensor products between a one-dimensional IR
$\Gamma\in\{A_1,A_2,B_1,B_2\}$ and the two-dimensional
representation $E$ satisfy
\begin{equation}
\Gamma\otimes E=E ,
\end{equation}
as equivalence classes of representations. To use the same $E$ matrices for every channel, a basis conversion is required:
\begin{equation}
    \widetilde{\bm f}^{(E)}=f^{(\Gamma)}T_\Gamma\bm f^{(E)}.
\end{equation}
For the Cartesian doublets of Appendix~\ref{app:ir-decomposition}, one may choose
\begin{align}
T_{A_1}&=\begin{pmatrix}1&0\\0&1\end{pmatrix}, &
T_{A_2}&=\begin{pmatrix}0&-1\\1&0\end{pmatrix},\notag\\
T_{B_1}&=\begin{pmatrix}1&0\\0&-1\end{pmatrix}, &
T_{B_2}&=\begin{pmatrix}0&1\\1&0\end{pmatrix}.
\end{align}
These matrices satisfy $T_\Gamma\chi_\Gamma(R)\mathcal A_E(R)=\mathcal A_E(R)T_\Gamma$, where $\chi_\Gamma$ is the one-dimensional character. Bare multiplication by $f^{(\Gamma)}$ generally gives an equivalent doublet in a different basis; it cannot be mixed componentwise with the original $E$ features without this conversion.

\begin{table}[t]
\begin{ruledtabular}
\begin{tabular}{c c c c c}
$\Gamma$ &
$C^\Gamma_{xx}$ &
$C^\Gamma_{yy}$ &
$C^\Gamma_{xy}$ &
$C^\Gamma_{yx}$ \\
\hline
$A_1$ & $+1$ & $+1$ & $0$  & $0$  \\
$A_2$ & $0$  & $0$  & $+1$ & $-1$ \\
$B_1$ & $+1$ & $-1$ & $0$  & $0$  \\
$B_2$ & $0$  & $0$  & $+1$ & $+1$ \\
\end{tabular}
\end{ruledtabular}
\caption{Nonzero Clebsch--Gordan coefficients $C^\Gamma_{ij}$ for
the decomposition $E\otimes E\rightarrow\Gamma$ in the point group
$D_4\simeq C_{4v}$, defined by
$f^{(\Gamma)}=\sum_{i,j=x,y}C^\Gamma_{ij}
f_i^{(E,1)}f_j^{(E,2)}$.}
\label{tab:cg-d4}
\end{table}

\begin{table}[b]
\begin{ruledtabular}
\begin{tabular}{c c c c c c}
 & $A_1$ & $A_2$ & $B_1$ & $B_2$ & $E$ \\
\hline
$A_1$ & $A_1$ & $A_2$ & $B_1$ & $B_2$ & $E$ \\
$A_2$ &       & $A_1$ & $B_2$ & $B_1$ & $E$ \\
$B_1$ &       &       & $A_1$ & $A_2$ & $E$ \\
$B_2$ &       &       &       & $A_1$ & $E$ \\
$E$   &       &       &       &       &
$A_1\oplus A_2\oplus B_1\oplus B_2$ \\
\end{tabular}
\end{ruledtabular}
\caption{Direct products of irreducible representations of the
$D_4$ point group.}
\label{table:D4-product}
\end{table}

The nontrivial Clebsch--Gordan decomposition involving two
$E$-type objects is
\begin{equation}
E\otimes E=A_1\oplus A_2\oplus B_1\oplus B_2 .
\end{equation}
Given two $E$-type objects
\begin{equation}
\bm f^{(E,1)}
=
\left(f_x^{(E,1)},f_y^{(E,1)}\right),
\qquad
\bm f^{(E,2)}
=
\left(f_x^{(E,2)},f_y^{(E,2)}\right),
\end{equation}
a convenient choice of symmetry-adapted bilinear combinations
transforming according to the one-dimensional IRs of $D_4$ is
\begin{eqnarray}
f^{(A_1)}
&=&
f_x^{(E,1)}f_x^{(E,2)}
+
f_y^{(E,1)}f_y^{(E,2)},\\
f^{(A_2)}
&=&
f_x^{(E,1)}f_y^{(E,2)}
-
f_y^{(E,1)}f_x^{(E,2)},\\
f^{(B_1)}
&=&
f_x^{(E,1)}f_x^{(E,2)}
-
f_y^{(E,1)}f_y^{(E,2)},\\
f^{(B_2)}
&=&
f_x^{(E,1)}f_y^{(E,2)}
+
f_y^{(E,1)}f_x^{(E,2)}.
\end{eqnarray}

These four combinations correspond, respectively, to the fully
symmetric scalar ($A_1$), the antisymmetric pseudoscalar ($A_2$),
and two anisotropic channels ($B_1$ and $B_2$), often associated
with $d_{x^2-y^2}$- and $d_{xy}$-like symmetry.

Equivalently, the CG structure of $E\otimes E$ may be written in
compact tensor form,
\begin{equation}
f^{(\Gamma)}
=
\sum_{i,j=x,y}
C^\Gamma_{ij}
f_i^{(E,1)}
f_j^{(E,2)},
\end{equation}
where the nonzero Clebsch--Gordan coefficients $C^\Gamma_{ij}$ are
listed in Table~\ref{tab:cg-d4}. They are given without normalization; multiplying each projection by $1/\sqrt{2}$ gives orthonormal CG coefficients. A consistent overall scale can be absorbed into the subsequent trainable weights. The antisymmetric $A_2$ contraction vanishes for two identical input doublets, but can be nonzero for distinct channels. For completeness, the direct-product rules for the $D_4$ IRs are summarized in Table~\ref{table:D4-product}.

\section{Equivariant neural network (ENN) model setup}
\label{app:enn-setup}

The cutoff $r_c=3.61$ gives 45 scalar inputs per site, organized into the IR multiplicities listed in Table~\ref{tab:enn-arch}. Each $E$ channel contains two components, whereas each one-dimensional channel contains one; thus the input multiplicities $(9,3,6,5,11)$ account for all 45 coordinates. The network has an input layer, three hidden layers, and a scalar output layer. The hidden layers contain 32, 8, and 4 copies of each IR, respectively. Their multiplicities describe learned channels rather than additional lattice sites, so changing these widths does not change the input neighborhood.

Between consecutive layers, direct features and selected tensor-product features form the symmetry-expanded sets in Table~\ref{tab:enn-arch}. Fully connected transformations act on channel indices within each IR, with the same weights applied to every component of a doublet. The tabulated expanded dimensions specify the retained feature sets; representation-equivalent copies must use consistent bases before this mixing. The hidden layers use the amplitude activation in Eq.~\eqref{eq:activation}, with $\sigma(u)=\max(0,u)$ (ReLU), and the final linear readout retains only $A_1$. Its scalar bias is allowed by symmetry. With shared doublet weights, the tabulated transformations contain 142,914 weights; 220 hidden amplitude biases and one output bias give the stated total of 143,135 parameters.

Training uses the force loss in Eq.~\eqref{eq:force-loss}. The data comprise 600 configurations on a $40\times40$ lattice, including random states, quasi-ordered CDW states, and intermediate thermal-quench configurations. The split assigns 400 configurations to training and 200 to testing, so every local environment from a given configuration belongs to the same subset. Local environments within a configuration can remain strongly correlated; their number should not be interpreted as a count of independent samples.

The model is implemented in PyTorch and optimized with Adam, using an initial learning rate of $10^{-3}$ and a batch size of 1. A scheduler adjusts the learning rate according to the training loss. Section~\ref{sec:holstein-benchmarks} tests both force errors and dynamical correlations, since instantaneous accuracy alone does not establish the accuracy of long-time evolution.

\bigskip

\noindent {\bf Data availability.} The machine learning code and sample training data used in this work can be found in the repository: \href{https://github.com/cherngroupUVA/Equivariant-FNN}{https://github.com/cherngroupUVA/Equivariant-FNN}. The code for dynamical simulation and more training data are available upon reasonable request.

\bibliography{ref}

\end{document}